\documentstyle[twocolumn,aps,floats]{revtex}

\begin{document}

\draft \tolerance = 10000

\setcounter{topnumber}{1}
\renewcommand{\topfraction}{0.9}
\renewcommand{\textfraction}{0.1}
\renewcommand{\floatpagefraction}{0.9}
\newcommand{\br}{{\bf r}}

%Fixing abstract in twocolumn mode
\twocolumn[\hsize\textwidth\columnwidth\hsize\csname
@twocolumnfalse\endcsname

\title{What Future Expects Humanity  After the  Demographic Transition Time? }
\author{L.Ya.Kobelev, L.L.Nugaeva  \\
 Department of  Physics, Urals State University \\ Lenina Ave., 51,
Ekaterinburg 620083, Russia  \\ E-mail: leonid.kobelev@usu.ru}
\maketitle

\begin{abstract}
The variant of phenomenological theory of humankind future existence after
time of demographic transition  based on treating the time of demographic
transition as a point of phase transition and taking into account an
appearing of the new phase of mankind  is proposed. The theory based on
physical phenomenological theories of phase transitions and classical
equations for system predatory-preys for two phases of mankind, take into
account assumption about a multifractal nature of the set of number of
people in temporal axis and contains control parameters. The theory
includes  scenario of destroying of existent now human population by new
phase of humanity and scenario of  old and new phases co-existence. In
particular cases when the new phase of mankind is absent the equations of
theory may be formulated as  equations of Kapitza , Foerster, Hoerner,
Kobelev and Nugaeva, Johansen and Sornette phenomenological theories of
growth of mankind.
\end{abstract}
 \vspace{1cm}
%Fixing abstract in twocolumn mode
]

\section {Introduction}
The problem of mankind population growth is  one of the global problems of
the mankind existence and  its prosperity in the  future. Will the
demographic explosion existing now at the mankind population of the whole
world stops and what will be after its stopping? Almost all the
phenomenological theories of future of mankind ( see
\cite{foer},\cite{hoer}, \cite{kap}, \cite{kob1}, \cite{sor}) include the
time  of demographic transition $t_{0}$. Before this time the growth of
mankind population is very large ( hyperbolic or exponential growth),
after this time the population growth may practically absent \cite{kap} or
may be regulated by driving parameters \cite{kob1} that govern the
development of mankind (such parameters are present in non-linear open
systems (see\cite{klim}). Thus the time point $t_{0}$ (the time of
demographic transition) is very essential for the future of mankind. What
is biological and  physical senses of the demographic transition time? The
mankind as the whole is a very large system and its behavior can be
described by methods of nonlinear dynamic, probability theory or
statistical theory of open system using the  methods of fractal geometry
(\cite{mad}, \cite{kob2}, or power laws \cite{sor}) . Now put the
question: how the demographical transition time may be treated from these
points of view? There are many different ways to answer on this question.
In this paper we consider the demographic transition  time as a point of a
phase transition, i.e. the point of transition of mankind in a new phase (
in physiological, psychological, behaviour and so on senses). What this
phase will be if it may exist? Will be the population of men of the new
phase friendly to the old phase of mankind or will they fight with men of
the old phase and force out them from their ecological niche? What
scenario may exist and be admissible in that case? It seems that if the
driving parameters that determine growth of the new phase ( thou it is
difficult now to select the main of them, it is the tusk for many research
groups ) will the same as now the new phase may diminish the old phase of
mankind very soon. We would remind that all phenomenological theories give
for the time $t_{0}$ the time in first part of twenty one century. If the
treating of time $t_{0}$ as a point of phase transition is correct and has
relation to existence of humanity , the embryos of the new phase must
appear ( as it well known from theory of phase transitions in physics)
before the time of demographic transition , i.e. in very near time. What
role they will play in the life of mankind society and what it will be
like? Will the new phase consist of international groups of islam
terrorist and the fight and wars spread to the whole world? Or it will
international groups of corrupted officials that for own prosperity govern
the whole world to its death? May be the groups of sexual minorities will
became the main part of mankind and diminish it ? Will the new phase
characteristics lay in psychological (i.e. behaviour) domain or it lay in
the physiological domain ( may be it is incurable genetics diseases or
mental disorders reasoned by increasing radiation in our world)? Are the
Russian revolution in 1917 year or the Germany fascist state in 1933 year
are the examples of appearing of such phases? May be the victory of
progressive mankind in the World War $II$ and collapse of socialist system
in Soviet Union are results only the fact that the demographic time
transition is not come? It seems, the search of answers on these questions
is the one of main tusks of governments and research demographical  gropes
of developed countries in near time.\\
\\ We consider in this paper only one way of mathematical analysis and
description of two phases of mankind (the old phase and the new phase) and
describe time behaviour of these phases by system of differential
equations that are  Volterra-Lotka equations for system of predatory-prey.
\section{Equations for describing two phases mankind time behaviour}
There are two simplest ways (physical and biological) for describing time
dependence of two phases (the old and the new) of mankind : \\
\\1. In physics the phenomenological theory of phase transitions was developed
by Landau (see \cite{lan}). Theory of phase transitions as appearing of
order parameters developed by Ginzburg-Landau \cite{gin}. These theory may
be used for describing time behaviour of two phases.  From physical point
of view the transition from the old phase of mankind to the new phase may:
a) take very short time (as transition of water to ice); b) take long time
and has intermediate state when two phases of mankind will co-existent (as
in the transition to superconductive state for superconductors of second
order);c) be characterized by stretched in time axes phase transition  as
in some ferroelectrics.  The differences of these cases determined by
characteristics of the  small embryos gropes of new phase and their
interaction with the old phase of mankind. If interaction between men of
old and new phases will attract the man of old phase to transition to the
new phase (we will call such interaction as "negative" ), the transition
time will be small and co-existent of two phases of mankind for the long
time period is impossible. Probably  the evidences of such transitions are
the above examples of Russian revolution of 1917 year and fascist Germany
state of 1933 year. If  result of interaction of men of the old phase with
gropes of men of the new phase will neutral, i.e. men of old phase will
not be attracted by embryos of new phase ( "positive" interaction), amount
of new the phase will increase and co-existence of new and old phases is
possible.  The time of coexistence determined by govern parameters and
selected mathematical models.
\\2. The biological point of view on behavior of mankind  allows
to use many mathematical models for describing of time behaviour of old
and new mankind phases. In this paper we see only simple predatory-prey
models where the old phase of mankind will be prey and the new phase will
be predatory.
 \section {Equations of Wolterra-Lotka for old and new phases
mankind population} Let us designate the old phase of mankind by
$N_{1}(t)$ and the new phase  of mankind by $N_{2}(t)$ where $t$ is
current time.  Equations of Volterra-Lotka type  is well researched. It is
well considered the time behaviour of the predatory and the prey
populations (including time behaviour populations near the point of
demographic transition $t_{0}$ where fluctuations play important role). So
we only write these equations and mark the main consequences of
mathematical theories.  Let for time $dt$ the old phase changes are
$cN_{1}^{\nu}dt - \gamma N_{1}N_{2}dt$ ( the second member is the result
of diminishing character of influences (interactions) between the old and
the new phases) and new phase changes are $-\delta N_{2}dt +
bN_{1}N_{2}dt$. The equations for $N_{1}(t)$ and $N_{2}(t)$ may be written
in this case as
\begin{equation}\label{1}
\frac{\partial}{\partial t}N_{1}(t)=cN_{1}^{\nu}(t) - \gamma
N_{1}(t)N_{2}(t)
\end{equation}
\begin{equation}\label{2}
\frac{\partial}{\partial t}N_{2}(t)=-\delta N_{2}(t)+ bN_{1}(t)N_{2}(t)
\end{equation}
where  $c, b, \delta, \gamma$ are constant and defined by selections of
considered models. These equations are Volterra  equations for $\nu=1$ and
are Lotka equations for $\nu=0$. The humanity phases $N_{1}$ and $N_{2}$
in (\ref{1}),(\ref{2}) have links and determine one another. \\ In the
case when populations of the old phase and the new phase are multifractal
sets on the time axes (see \cite{kob1}) equations (\ref{1}) and (\ref{2})
read
\begin{equation}\label{3}
D_{0,t}^{\beta}N_{1}(t)=cN_{1}^{\nu}(t) - \gamma N_{1}(t)N_{2}(t)
\end{equation}
\begin{equation}\label{4}
  D_{0,t}^{\mu}N_{2}(t)=-\delta N_{2}(t)+ bN_{1}(t)N_{2}(t)
\end{equation}
where $D_{0,t}^{\beta}$ and $D_{0,t}^{\mu}$ are Riemann-Liouville
fractional derivatives or generalized Riemann-Liouville derivatives
(see\cite{kob1})and $\beta$ and $\mu$ are fractional numbers or functions
of govern parameters. \\ The equations (\ref{1}) and (\ref{2}) may be
considered as the base equations of rude bioligical models for describing
future population of mankind if there are exist two phases of mankind and
sets of $N_{1}$ and $N_{2}$ are sets of population of old and new phases .
As well known the equations of such type ( type of Volterra-Lotka
equations) may be generalized and wrote in more general form ( form of
Kolmogorov-Fokker-Plank equations for probability density
$w(N_{1},N_{2},t)$ to find in moment $t$ the populations $N_{1}$ and
$N_{2}$ in the considered  system \cite{kost}, \cite{vas}, or may be
considered  the case when influences $N_{2}$ on behaviors of $N_{1}$
included only at the bound time intervals and so on). These generalized
equations are wide spread and give many known models have used not only in
biology (\cite{mur}), but in economics (\cite{lor}), non linear physics,
theoretical chemistry  and so on. In the next part of paper we write more
detailed description of results of  interactions between  two mankind
phases $N_{1}{1}$ and $N_{2}$ based on well known researches of equations
of such type. For case when in above equations there are only one phase
$N_{1}$, and $\nu=0$ and coefficient $c$ is function of time and $t_{0}$
($c=c(t,t_{0})$ it is not difficult to chose $c(t,t_{0})$ in the simple
forms and as special cases receive the base equations of work
\cite{kap},\cite{kob1}, \cite{sor}.

\section{How will interact  old and  new mankind phases?}

The mathematical research of solutions characteristics of equations
(\ref{1})-(\ref{2}) or more general equations taking into account the
fluctuations is well worked out (see for example \cite{vas}. As was
pointed out these equations may be written  for mean values of $N_{1}$ and
$N_{2}$ or in the form of govern equations for probability
$w(N_{1},N_{2},t)$ to find the system in the moment $t$ with values
$N_{1}$ and $N_{2}$, or in the form of Kolmogorov- Fokker-Plank equations
that are consequences of governor equations.  The influences of noises and
fluctuations( depending at $c, b, \delta, \gamma$ ) on the behaviour of
system may be taken into account too. More complicated equations may be
written for case when the phase $N_{2}$ may part of her time have "a rest
time" and in that rest time do not interact with the phase $N_{1}$. So we
omit the process of receiving of solutions for different systems of
equations describing interaction between old-new phases and only describe
main results:\\ 1. Possibility of co-existence between old and new phases
with periodical changing of value of $N_{1}$ and $N_{2}$; \\2. Possibility
for diminishing of one phase by other phase that depends at the initial
values;\\3. Possibility for calculation of mean values of $N_{1}$ and
$N_{2}$ in one of system states;\\4. Possibility for calculation of time
existence $t_{ex}$ for old or new phases if  take into account by
including in the main equations (or changing the main equation by its
Fokker-Plank probability analogies equations) external or internal noises
( additive or multiplicative);\\5. Possibility of description of
fluctuation characteristics of kinetic variables near bifurcation
points;\\6. Possibility  of researching of noises role in losing stability
of stationary state or in different self-excited oscillations states;\\7.
Possibility of determination of conditions for stochastic regimes
appearing in system;
\\ 8. Possibility of governing by processes of fighting or co-existence between old
and new phases;\\9. Possibility of using more complicated models and
equations including fractal characteristics of mankind population
distribution on the time axis.\\10. Possibility of selection of different
variants for govern parameters based on experimental sociological dates
and so on. \\Thus the predatory-prey models and its equations give rich
possibilities for researching of the behavior of two phases of mankind.
Has this assumption about  appearing of two mankind phases  relation to
problem of future propagation of mankind? The future propagation of
mankind will gives answer on this question, but if it assumption is
correct it is time now to find the new phase (or phases) and determine (or
make attempts to determine) is it one or other phases useful  for future
of mankind or not useful.
\section {Conclusions}
As one of advantages to use  the equations (\ref{1})-(\ref{2}) for
describing demographic problems (with some of them the mankind already has
confronted now and may be these problems are consequences  of appearing
the embryos of new phase of mankind ) we shall stress an opportunity of
insert and account in the theory many factors (such as, incurable
illnesses, natural cataclysmic etc.) defining future of mankind as a
result of influences the control parameters included in the  govern
parameters of predatory-prey models and fractional dimensions .\\ There
are three main scenario of models based on equations type of
(\ref{1})-(\ref{2}):\\a) Two phases of mankind co-existent during long
time with periodically changes its values and with mean values
$N_{1}=\delta b^{-1}$ and $N_{2}=c\gamma^{-1}$ conserves in bound limits
if population fluctuations may be omitted;
\\b) Mankind population  lost its stability as the result of fluctuation
influences if the equations of Lotta type used;\\c)The old mankind phase
has  bound time of existence and this time may by calculated if parameters
of theory are defined .
\\The main purpose of this paper was to analyze possibility of
introducing of two mankind's population phases after demographic
transition time. Driving parameters included in the considered models
(equations of Landau-Ginzburg time theory or equations of predatory-prey
type) of the phenomenological theories of the mankind population in the
frame of physical Landau or Ginzburg-Landau models or biological
predatory-prey models. The  predatory-prey type models was chosen asan
example. The change of number of mankind (described in the framework of
phenomenological theories of the population) can be adjusted by such
choice of control parameters  with  concrete contents of dependencies of
them  at which the mankind will has stability time and population will
grow so slowly, that overpopulation and the problems connected with
fighting between two mankind  phases will do not arise in the foreseen
future. Stress now as was done in \cite{sor} that after demographic
transition time also the type of economical development of humankind may
be changed .

\end{document}